\documentclass[aps,prb,onecolumn,amsmath,amssymb,12pt]{revtex4-1}  

\usepackage{graphicx}  
\usepackage{dcolumn}   
\usepackage{bm}        
\usepackage{amssymb}   
\usepackage{subfigure}   
\usepackage{epstopdf}
\usepackage[version=3]{mhchem} 
\usepackage[T1]{fontenc}       

\begin{document}

\title{ Broadband suppression of backscattering at optical frequencies using low permittivity dielectric spheres}

 \email{m.abdelrahman@fresnel.fr}

\author{M. Ismail Abdelrahman$^{* 1,2}$, C. Rockstuhl$^{1,3}$, and I. Fernandez-Corbaton$^{3}$. }
\affiliation{%
 $^{1}$ Institute of Theoretical Solid State Physics, Karlsruhe Institute of Technology, Karlsruhe, Germany \\
$^{2}$ Aix Marseille Univ, CNRS, Centrale Marseille, Institut Fresnel, Marseille, France \\
$^{3}$ Institute of Nanotechnology, Karlsruhe Institute of Technology, Karlsruhe, Germany
}%

\date{\today}

\begin{abstract}
The exact suppression of backscattering from rotationally symmetric objects requires dual symmetric materials where $\epsilon_\textrm{r}=\mu_\textrm{r}$. This prevents their design at many frequency bands, including the optical one, because magnetic materials are not available. Electromagnetically small non-magnetic spheres of large permittivity offer an alternative. They can be tailored to exhibit balanced electric and magnetic dipole polarizabilities, which result in approximate zero backscattering. In this case, the effect is inherently narrowband. Here, we put forward a different alternative that allows broadband functionality: Electromagnetically large spheres made from low permittivity materials. The effect occurs in a parameter regime that approaches the trivial $\epsilon_\textrm{r}\rightarrow\mu_\textrm{r}=1$ case, where approximate duality is met in a weakly wavelength dependence fashion. Despite the low permittivity, the overall scattering response of the spheres is still significant. Radiation patterns from these spheres are shown to be highly directive across an octave spanning band. The effect is analytically and numerically shown using the Mie coefficients.
\end{abstract}

\maketitle

The scattering of light upon interacting with matter is a central problem in electromagnetism which is relevant in many branches of physics such as nuclear physics, astrophysics, and spectroscopy. The theory of light scattering from spheres was developed more than a century ago by Gustave Mie\cite{mie1908beitrage}. Due to the intricate nature of the multiple phenomena taking place in this interaction process, the exploration of different regimes continues to disclose surprising new effects that have a profound impact on a wide range of applications\cite{alu2008tuning, knoll2000enhanced, moreno2013analysis, sarmecanic1997constraints}. 
A referential example is the desire to suppress the backscattering from spheres that could find use, e.g. in light management structures used in photovoltaic devices\cite{ferry2010design,  akimov2010nanoparticle,  zhang2015dielectric} and laser tractor beams  (negative optical force)\cite{chen2011optical}. In 1983 Kerker {\it et al.}\cite{kerker1983electromagnetic} showed that it is possible to achieve zero backscattering (ZBS) in a direction opposite to the illumination using spheres made from a dual material\cite{FerCor2012p}, i.e. a material with identical electric permittivity and magnetic permeability $\epsilon_\textrm{r}=\mu_\textrm{r}$. Electromagnetic duality implies the equal complex amplitude excitation of electric and magnetic multipolar moments inside the sphere. Their scattered fields interfere destructively in the backward direction. The condition of spherical symmetry can be relaxed to cylindrical symmetry\cite{Zambrana2013} and even to discrete rotational symmetries\cite{FerCor2013c}. However, no natural material with magnetic properties exists at optical frequencies. In other words, all materials exhibit a  permeability of $\mu_\textrm{r}=1$. Thus, Kerker's method can not be applied at optical frequencies. While metamaterials might be an avenue to solve this problem, there has been no report of a dual metamaterial so far. Moreover, the absorption inherent to the resonant inclusions needed to achieve an effective permeability larger than one would likely spoil the entire effect. 

Nevertheless, it has been shown, both experimentally\cite{kuznetsov2011magnetic} and theoretically\cite{garcia2011strong}, that it is possible to excite a notable magnetic dipole moment at optical frequencies in  nanospheres made from a high permittivity dielectric material where intrinsic losses can be negligible. When the radius and permittivity of small spheres are carefully selected, it is possible to achieve duality in the dipolar approximation\cite{zambrana2013dual}, albeit for a very narrow spectral region. This leads to the possibility of approaching ZBS in experiments, as shown in Ref. \citenum{geffrin2012magnetic} in the GHz band, and Ref. \citenum{fu2013directional} in the optical band.  Dielectric spheres of moderate permittivity can also exhibit forward scattering  at their  scattering peak in the visible region\cite{zhang2015colloidal}. Zero backscattering has also been experimentally approached in the optical band using electromagnetically small cylindrical objects\cite{person2013demonstration}. Arrays of silicon disks have  been used to design near-unity transmission Huygens's surfaces  under normal illumination in an extended spectral region \cite{staude2013tailoring,decker2015high}. As the size of the objects grows, higher order multipoles become significant and, while they can be  used for ZBS interference in some cases\cite{alaee2015generalized,li2014broadband}, they are a source of duality breaking and ZBS degradation in homogeneous spheres. This is illustrated in Fig.~1
. Consequently, achieving a broadband ZBS from spheres at optical frequencies remains elusive.

\begin{figure}
    \centering
    \includegraphics[width=0.7\linewidth, keepaspectratio]{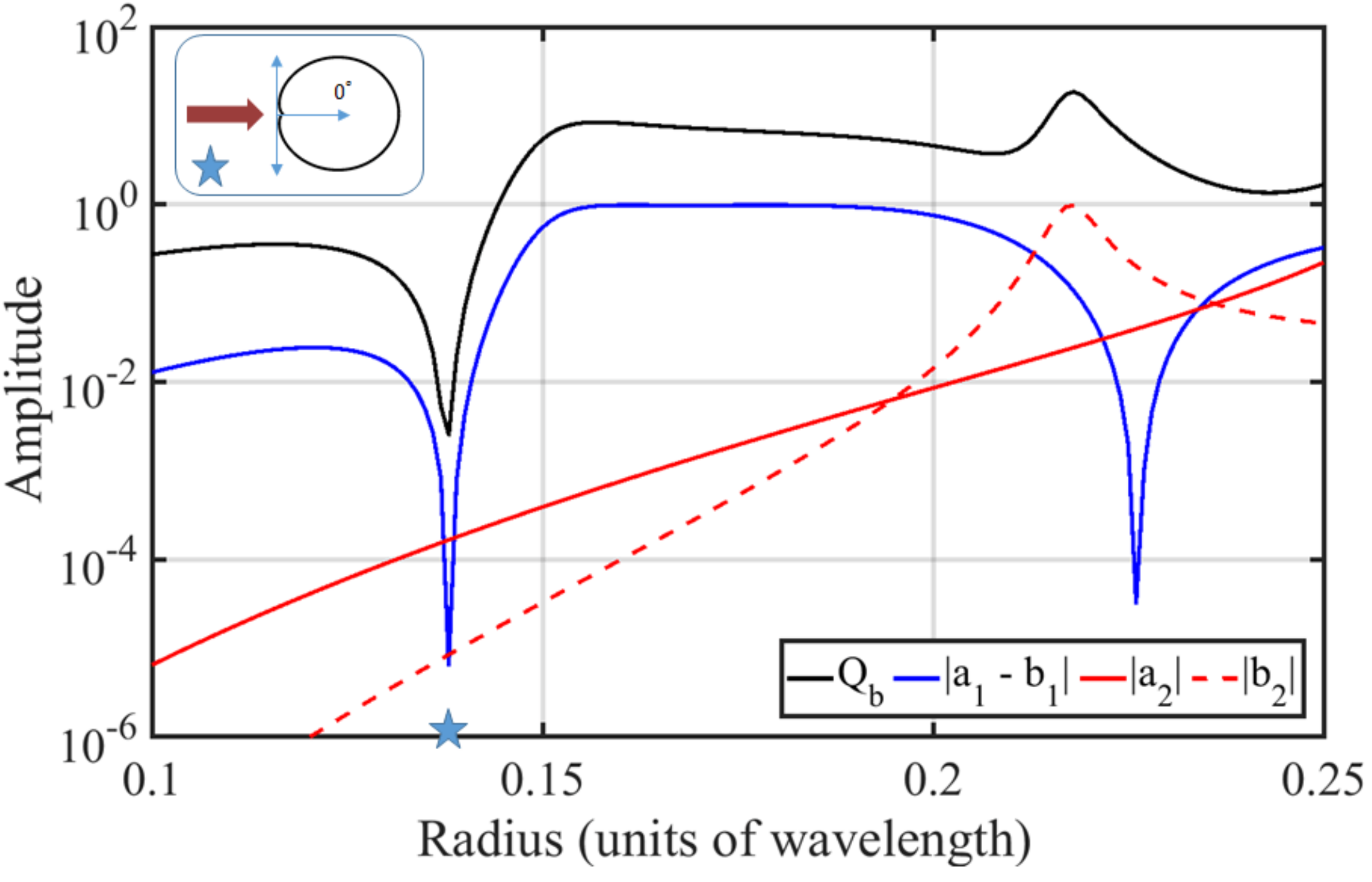}
	\caption{Backscattering efficiency $Q_\textrm{b}$  of a nonmagnetic sphere with permittivity $\epsilon_\textrm{r}=10$ as a function of radius in units of wavelength $\lambda_0$. At $r =0.138\, \lambda_0$, the electric and magnetic dipoles are balanced ($a_1=b_1$), and the scattering pattern resembles a Huygens source, as shown in the top-left corner. The figure shows another point where $a_1=b_1$ at  $r =0.226\, \lambda_0$, however, this larger sphere exhibits  considerable backscattering due to the notable excitation of high order multipoles. }
    \label{fig:1ZBSdipole}
\end{figure}

The basic problem can be formulated quite simply: Can we design a broadband strongly scattering object with a heavily reduced backscattering? In this article, we show that the answer is positive for electromagnetically large dielectric spheres made from low permittivity materials that, if carefully designed, exhibit both ZBS and strong interaction in an extended spectral domain. We explain the underlying physics by analytical means and corroborate our findings with calculations based on Mie coefficients. To quantify the relevant features and to analytically motivate the chosen path, we consider the backscattering efficiency $Q_\textrm{b}$\cite{kerker2016scattering} as a parameter to judge the backscattering characteristics of spheres of arbitrary sizes and study its analytical form for a sphere with a permittivity close to the duality point $\epsilon_\textrm{r}=\mu_\textrm{r}$. The backscattering efficiency represents the ratio of the intensity scattered in the backward direction ($\theta=180^{\circ}$) to the intensity of the illumination. It is directly related to the electromagnetic duality of the sphere: Zero backscattering is a property of objects that are dual symmetric and, additionally, have (discrete) rotational symmetry\cite{Zambrana2013,FerCor2013c}.  Mie theory provides an analytical expression for the backscattering efficiency  as\cite{kerker1983electromagnetic,bohren1983absorption}
 \begin{equation}  \label{eq:Qb}
Q_\textrm{b} =  \frac{1}{ x^2}\; {\left| \sum_{n=1}^{\infty} (2n+1)\;(-1)^\textrm{n}\;(a_\textrm{n}-b_\textrm{n}) \right|}^2,
\end{equation}  
where $n$ is the order of the multipole expansion  associated with electric and magnetic Mie coefficients, $a_\textrm{n}$ and $b_\textrm{n}$, respectively. The size parameter  $x = 2 \pi r / \lambda_\textrm{0}$  represents the ratio between the sphere radius $r$  and the excitation wavelength $\lambda_\textrm{0}$. The backscattering efficiency depends on the terms $a_\textrm{n}-b_\textrm{n}$, which represent both the destructive interference between the electric and magnetic response of the sphere in the backward scattering direction, and also the breaking of electromagnetic duality by the sphere. For spheres, electromagnetic duality $\epsilon_\textrm{r} = \mu_\textrm{r}$ is equivalent to having identical electrical and magnetic responses $a_\textrm{n} = b_\textrm{n}$ for all $n$, thus zero backscattering $Q_\textrm{b}=0$. 

Another important quantity to consider is the scattering efficiency $Q_\textrm{sca}$, which is directly related to the total scattering power. Achieving a vanishing backscattering would be pointless if no light is scattered at all. This quantity measures the strength of the interaction between the illumination and the sphere. The scattering efficiency is given by\cite{kerker2016scattering}
\begin{equation}
	\label{eq:qsca}
Q_\textrm{sca} =  \frac{2}{x^2}\;  \sum_{n=1}^{\infty} (2n+1)\;( \left|a_\textrm{n} \right|^2 + \left|b_\textrm{n} \right|^2 ).
\end{equation}
In a nutshell, the broadband effect is due to approaching but not quite reaching the trivial $\epsilon_\textrm{r}\rightarrow\mu_\textrm{r}=1$ case. In this limit, the duality breaking terms which are responsible for backscattering, i.e. the difference between electric and magnetic Mie coefficients $(a_n-b_n)$, tend to zero, and so does hence the backscattering power [Eq. (\ref{eq:Qb})]. Crucially, we analytically show that the sensitivity of the backscattering power with respect to the electromagnetic size of the sphere tends to zero as well. This is the reason for the broadband character of the effect. In this regime, the electric and magnetic Mie coefficients of the spheres are approximately equal for all multipolar orders, which means that electromagnetic duality is approximately met even though nonmagnetic materials are considered. Despite the rather low permittivity, the proposed spheres scatter strongly due to their rather large diameter, roughly between 1.55 and 3.1 times the wavelength.

\section*{Results}

\subsection{Theoretical concept: Backscattering behavior  near duality.}
\label{sec:theory}

\def\px{\psi_\textrm{n}(x)}
\def\pmx{\psi_\textrm{n}(mx)}
\def\dpx{\psi'_\textrm{n}(x)}
\def\ddpx{\psi_\textrm{n}^{''}(x)}
\def\dpmx{\psi'_\textrm{n}(mx)}
\def\ddpmx{\psi_\textrm{n}^{''}(mx)}
\def\zx{\zeta_\textrm{n}^{(1)}(x)}
\def\dzx{\zeta_\textrm{n}^{(1)'}(x)}
\def\ddzx{\zeta_\textrm{n}^{(1)''}(x)}

The main idea of  this article is to employ the near-duality region, where $\epsilon_\textrm{r} \approx \mu_\textrm{r}$, to achieve a broadband suppression of backscattering, particularly for nonmagnetic spheres of free space permeability $\mu_\textrm{r}=1$. We will now analyze the sensitivity of the backscattering to changes in wavelength in this parameter regime. To this end, we employ the analytic expressions of the Mie coefficients. 

The Mie coefficients of a sphere with arbitrary electric permittvity $\epsilon_\textrm{r}$ and magnetic permeability $\mu_\textrm{r}$  are given by\cite{bohren1983absorption,grainger2004calculation}
\begin{equation}
	\label{eq:an}
a_\textrm{n}(x,m) = \frac{\px \dpmx - \tilde{m} \dpx \pmx } {\zx \dpmx - \tilde{m} \dzx \pmx},
\end{equation}
\begin{equation}
	\label{eq:bn}
b_\textrm{n}(x,m) = \frac{\dpx \pmx - \tilde{m} \px \dpmx } {\dzx \pmx - \tilde{m} \zx \dpmx},
\end{equation}
where $m$ is the  refractive index of the sphere relative to the ambient medium, assumed to be free space, and $\tilde{m} = m / \mu_\textrm{r}$. The prime represents the derivative of the function with respect to its argument. The functions $\psi_\textrm{n}(x)$ and $\zeta_\textrm{n}^{(1)}(x)$ are Riccati-Bessel functions defined in terms of the spherical Bessel and Hankel functions of the first kind, $j_\textrm{n}(x)$ and $h_\textrm{n}^{(1)}(x)$\cite{abramowitz1964handbook},
\begin{equation}
\px= x\;j_\textrm{n}(x), \; \; \zx= x\;h_\textrm{n}^{(1)}(x).
\end{equation}
The sensitivity ($V$) of the backscattering with respect to wavelength changes (or more generally the size parameter $x$) while approaching the duality point, i.e. when $m= \sqrt{\epsilon_\textrm{r}\;\mu_\textrm{r}} \to \mu_\textrm{r}$, thus $ \tilde{m} \to 1$, can be defined as
\begin{equation} \label{eq:limdQb}
V = \lim_{\tilde{m} \to 1}\; \frac{\partial Q_\textrm{b}(x)}{\partial x}. 
\end{equation}
Using that the differentiation of a square modulus of a complex function is given by the identity 
\begin{equation}
\frac{ \partial {|f(x)|}^2}{\partial x} = 2 \left[ \Re\{f(x)\} \Re\{f'(x)\} + \Im\{f(x)\} \Im\{f'(x)\} \right], 
\end{equation} 
where $\Re\{\cdot\}$ and $\Im\{\cdot\}$ denote the real and imaginary parts of the arguments, respectively, we expand the derivative of the backscattering efficiency in Eq. (\ref{eq:limdQb})
\begin{equation} \label{eq:dQb}
\begin{split}
 \frac{\partial Q_\textrm{b}(x)}{\partial x} =   \frac{2}{ x^2} \left\{   \sum_{n=1}^{\infty} (2n+1)\;(-1)^\textrm{n}\;\Re{\big(a_\textrm{n}-b_\textrm{n}\big)} \;  \sum_{l=1}^{\infty} (2l+1)\;(-1)^\textrm{l}\;\Re{\left[\frac{\partial(a_\textrm{l}-b_\textrm{l})}{\partial x}\right]}\right. \\ + \left.\sum_{n=1}^{\infty} (2n+1)\;(-1)^\textrm{n}\;\Im{\big(a_\textrm{n}-b_\textrm{n}\big)} \; \sum_{l=1}^{\infty} (2l+1)\;(-1)^\textrm{l}\;\Im{\left[\frac{\partial(a_\textrm{l}-b_\textrm{l})}{\partial x}\right]} \right\} \\  - \frac{2}{x^3}   {\left| \sum_{n=1}^{\infty} (2n+1)\;(-1)^\textrm{n}\;(a_\textrm{n}-b_\textrm{n}) \right|}^2,
\end{split}
\end{equation}
where $n$ and $l$ are positive integers. Accordingly, near the duality point, the backscattering sensitivity $V$ defined in Eq. (\ref{eq:limdQb}) depends on  the following limits,
\begin{equation} \label{eq:lim1}
\lim_{\tilde{m} \to 1}\;(a_\textrm{n}-b_\textrm{n})\; (a_\textrm{l}-b_\textrm{l}),
\end{equation}
\begin{equation} \label{eq:lim2}
\lim_{\tilde{m} \to 1}\; (a_\textrm{n}-b_\textrm{n}) \; \frac{\partial(a_\textrm{l}-b_\textrm{l})}{\partial x}.
\end{equation}
Using  properties of limits, the limit of a product of functions is equal to  the product of the limits of each function individually, if they both exist. The latter assumption is valid since Mie coefficients are bounded and continuous in the range $[0,1]$. Therefore, the backscattering sensitivity near the duality point vanishes if the terms $a_\textrm{n}-b_\textrm{n}$ tends to zero near the duality point. The terms $a_\textrm{n}-b_\textrm{n}$ turn out to be,
\begin{equation} \label{eq:qnbn}
(a_\textrm{n}-b_\textrm{n}) = (1-\tilde{m}^2) \frac{ \px \dpmx \dzx \pmx - \dpx \pmx \zx \dpmx  }{ \left[\zx \dpmx - \tilde{m} \dzx \pmx \right]  \left[\dzx \pmx - \tilde{m} \zx \dpmx \right]}.
\end{equation}
We note that the denominator of $a_\textrm{n}-b_\textrm{n}$ in Eq. (\ref{eq:qnbn}) can't be zero since it is the product of the denominators of $a_\textrm{n}$ and $b_\textrm{n}$ [see Eqs. (\ref{eq:an}) and (\ref{eq:bn})], and the Mie coefficients are bounded to the range $[0,1]$. Therefore, the value of Eq. (\ref{eq:qnbn}) approaches zero as $\tilde{m}$ approaches one, independently of both the multipolar order and the size parameter. Thus, the backscattering sensitivity near the duality point  [see Eq. (\ref{eq:limdQb})] tends to zero ($V \to 0$). This is the most important finding of our article. It identifies a regime that features the wavelength independent vanishing of the backscattering contributions for all multipolar orders $n$. This suggests a wavelength independent behavior, which leads to broadband backscattering suppression. This effect resembles the Kerker condition of ZBS but for nonmagnetic spheres made from low permittivity materials $\epsilon_\textrm{r} \approx (\mu_\textrm{r}=1)$. Indeed, this is a regime of approximately duality: $a_\textrm{n} \approx b_\textrm{n}$ for all $n$.
\subsection{Evaluation based on Mie coefficients: Parameter regime for broadband ZBS.}

\begin{figure}
    \centering
    \includegraphics[width=0.6\linewidth, keepaspectratio]{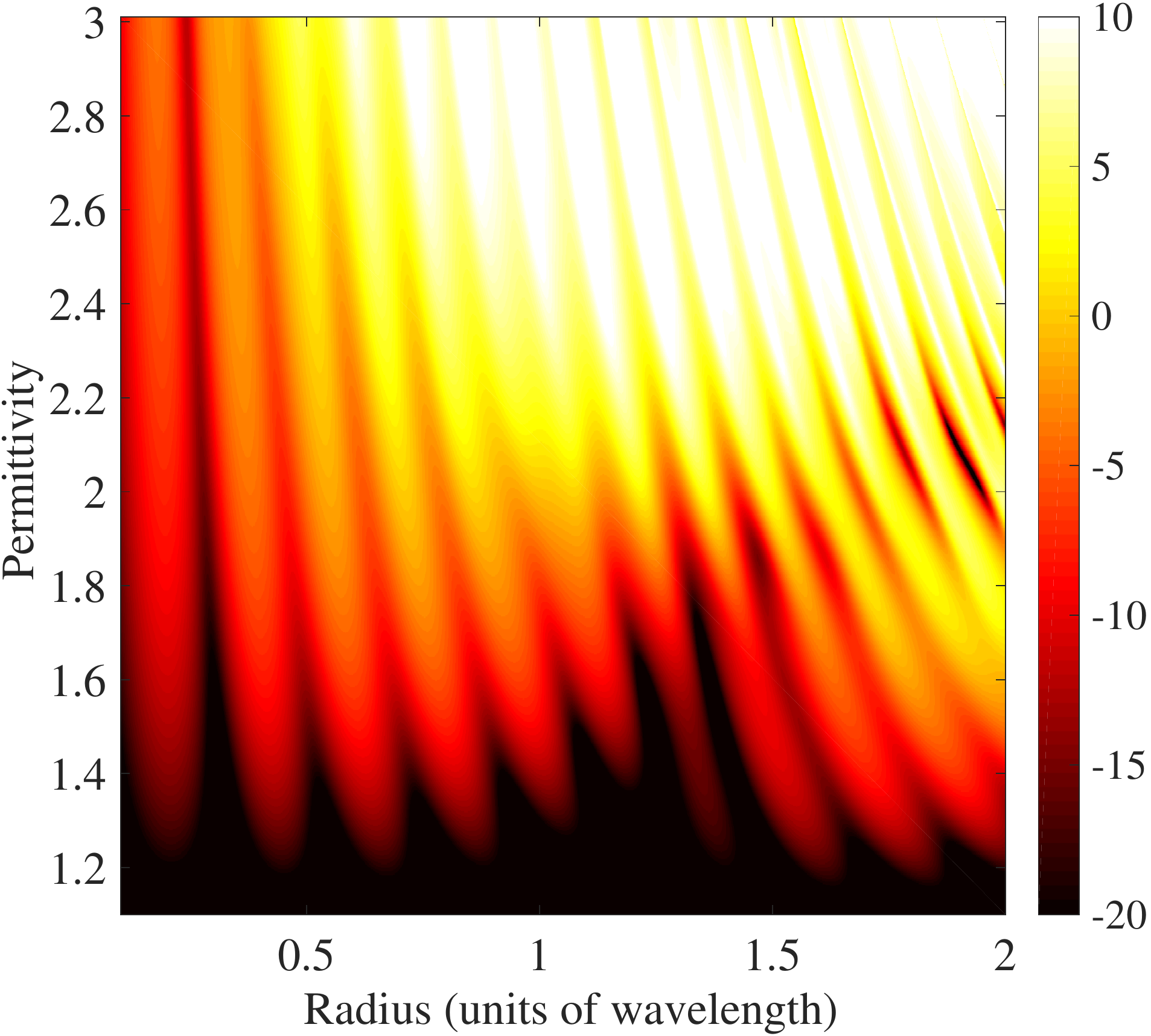}
	\caption{Backscattering efficiency $Q_\textrm{b}$  [dB scale]  of nonmagnetic spheres for a wide range of permittivities and radii. The black color indicates the regions of ZBS. Approaching the duality point $\epsilon_\textrm{r}\rightarrow \mu_\textrm{r}=1$ leads to a broadband suppression of backscattering. }
    \label{fig:2Qb}
\end{figure}

\begin{figure}
    \centering
{\includegraphics[width=0.6\linewidth]{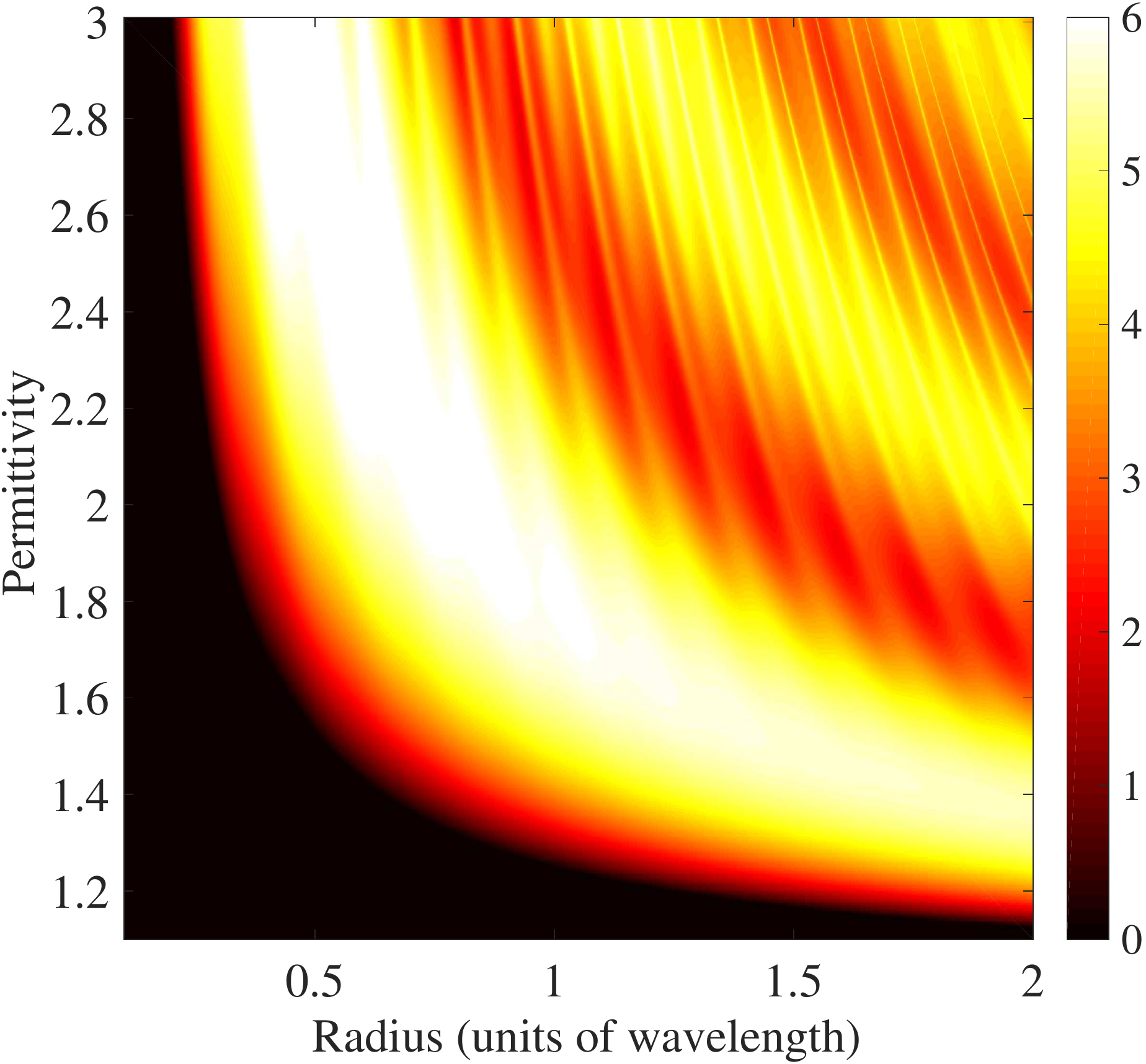}}	
\caption{ Scattering efficiency $Q_\textrm{sca}$  [dB scale]  of nonmagnetic spheres for a wide range of permittivities and radii. The white color indicates the regions of interest that show maximum interaction between the light and the sphere.  } \label{fig:3Qsca}
\end{figure}

\begin{figure}
    \centering
{\includegraphics[width=0.6\linewidth]{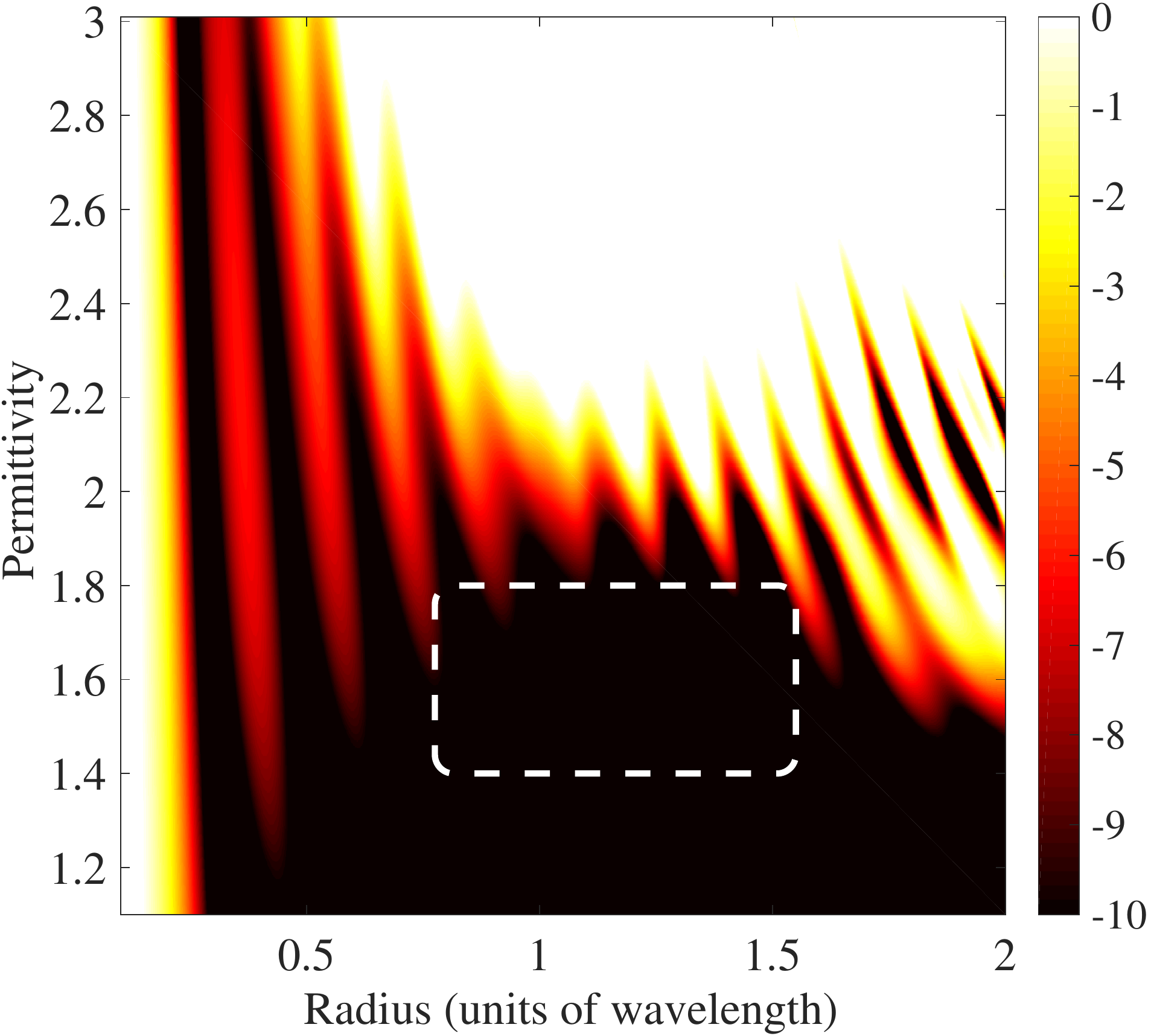}}	\caption{  Backscattering ratio $R_\textrm{b} = Q_\textrm{b}/Q_\textrm{sca}$  [dB scale]  of nonmagnetic spheres for a wide range of permittivities and radii. The dotted box  indicates a broadband ZBS region that shows both a low backscattering ratio $R_\textrm{b}$  and a considerable scattering efficiency $Q_\textrm{sca}$.  } \label{fig:4QbQsca}
\end{figure}

The results shown in the previous section suggest a parameter regime of low permittivity for broadband ZBS. Figure~2 
 shows the backscattering efficiency $Q_\textrm{b}$ of a nonmagnetic sphere as a function of its radius and material permittivity.  The dark regions in Fig.~2 represent parameter regions of vanishing backscattering.  As predicted in the previous Section, these ZBS regions get spectrally broadened for low values of the permittivity, i.e. in the near-duality region. Figure~3 shows $Q_\textrm{sca}$ of the same nonmagnetic sphere again as a function of its radius and material permittivity. The white color indicates parameter regions of a considerable scattering efficiency, which is a desired feature. 

Eventually, we are interested in minimizing the ratio of the backscattering efficiency to the scattering efficiency, namely, the backscattering ratio $R_\textrm{b} = Q_\textrm{b} / Q_\textrm{sca}$, given a large enough value of  the scattering efficiency $ Q_\textrm{sca}$. Figure~4 shows the backscattering ratio $R_\textrm{b}$ of the same nonmagnetic sphere as a function of its radius and material permittivity. The dotted box in Fig.~4  indicates an octave-wide spectral ZBS region for a radius  between 0.775 and 1.55 in units of the wavelength, that shows both a backscattering ratio $R_\textrm{b}$ below $10\%$ ($-10$ dB) and a scattering efficiency $Q_\textrm{sca}$ above $2$ ($3$ dB). We conclude that broadband backscattering suppression combined with a considerable scattering efficiency is possible using a wavelength-sized sphere made from a low permittivity material. An example for such material is Teflon\cite{yang2008optical} that exhibits a permittivity  around $1.7$ across the visible and near-IR spectra.

\section*{Discussion}


\begin{figure}
    \centering
{\includegraphics[width=1\linewidth]{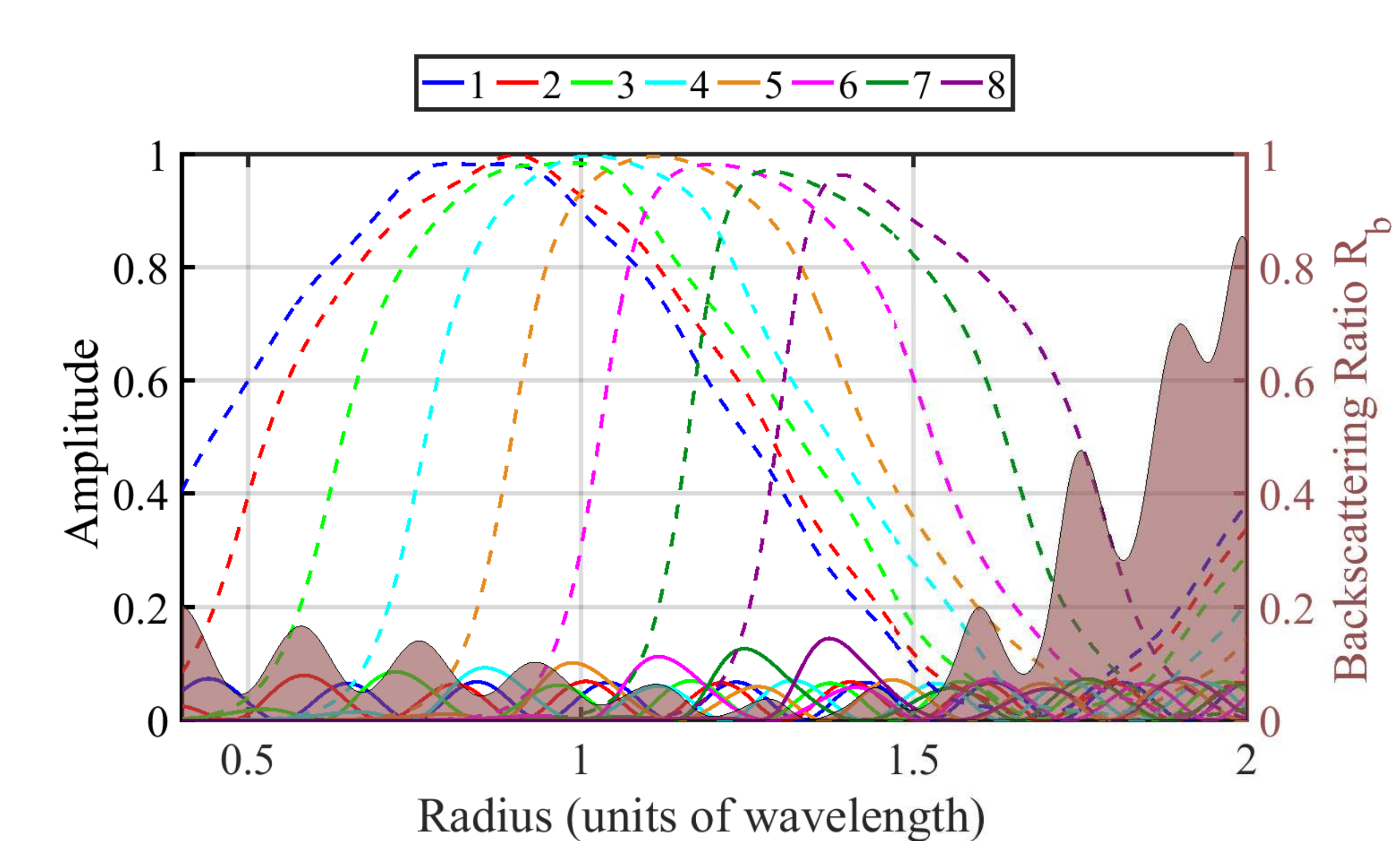}}	
\caption{Solid area: Backscattering ratio $R_\textrm{b}$. Dashed lines: Scattering efficiency of each multipolar order $\frac{1}{2}\left(|a_n|^2+|b_n|^2\right)$, for $n=1\ldots 8$. Solid lines: Duality breaking due to each multipolar order $|a_n-b_n|^2$.}
 \label{fig:5Mieexample} 
\end{figure}
%
In order to further investigate this broadband effect, we use the Mie coefficients to analyze a nonmagnetic sphere of permittivity $1.7$ for the radii corresponding to the region indicated in Fig.~4.  Figure 5 displays the quantities
\begin{equation}
	\label{eq:sd}
	\frac{1}{2}\left(|a_n|^2+|b_n|^2\right),\text{ and }|a_n-b_n|^2
\end{equation}
up to the eighth multipolar order. The first quantity in Eq. (\ref{eq:sd}) is related to the scattering due to the $n$-th multipolar order (see Eq. \ref{eq:qsca}). The second quantity is the duality breaking due to the $n$-th multipolar order. It is easy to see that $|a_n-b_n|^2$ are the helicity changing terms that determine the numerator of the duality breaking measure introduced in Eq. 2 of Ref. \citenum{FerCor2015}. According to Eq. (\ref{eq:Qb}), these terms are related to ZBS in the sense that when $a_n=b_n$, the $n-$th multipolar order does not contribute to the backward scattering.

Figure 5 shows low duality breaking for all multipolar orders across a wide range. This matches the analytical result obtained previously: Broadband backscattering suppression due to wavelength independent approximate duality $a_\textrm{n}\approx b_\textrm{n}$. Additionally, the scattering due to each multipolar term reaches its maximum (resonant) value inside the 0.775-1.55 band, and many have simultaneously large values in this region. This last feature is consistent with wide resonances characteristic of low refractive indexes.


\begin{figure}
\label{fig:6testcasePattern}
    \centering
 \includegraphics[width=1\linewidth, keepaspectratio]{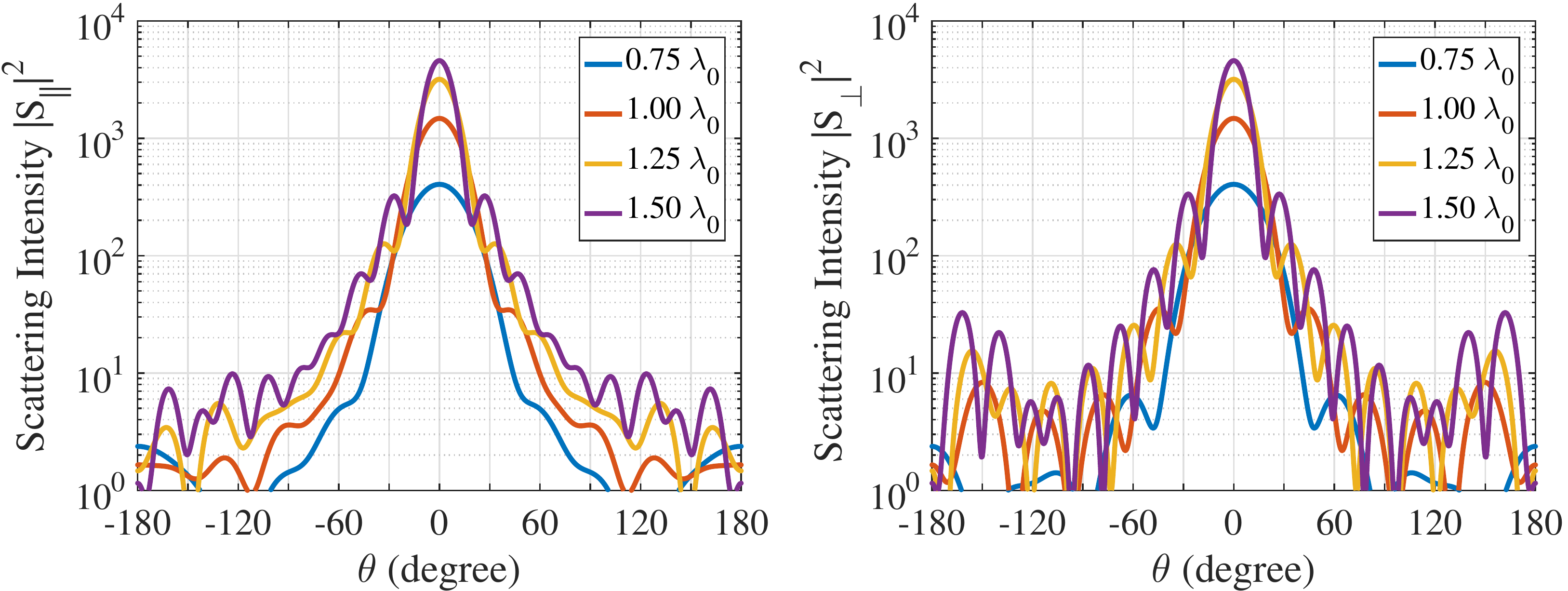}
    \caption{Angular distribution of the scattering intensities  (left) in the scattering plane, and (right) perpendicular to the scattering plane of a nonmagnetic sphere of  permittivity $1.7$ for different radii. The scattering efficiency  $Q_\textrm{sca}$ values are $3.07$, $3.8$, $3.63$, and $2.87$ for radius/wavelength ratios of $0.75$, $1.0$, $1.25$, and $1.5$,  respectively. }    
 \end{figure}
The proposed sphere of permittivity $1.7$ exhibits a dominant directional scattering pattern in the forward direction for the considered octave frequency band, as seen in Fig.~6.  Figure~6 shows the angular distribution of the  scattering  intensities, as a function of radius, both in and perpendicular to the scattering plane, $|S_\parallel|^2$ and $|S_{\bot}|^2$, respectively.  $\theta$ is the  angle in the scattering plane that contains the source, the sphere, and the observer. The scattering pattern can be judged using the asymmetry parameter $g$, which indicates the value of the average cosine of the distribution of scattering intensity\cite{bohren1983absorption} 
\begin{equation}
g = < u > = \dfrac{1}{x^2 \; Q_\textrm{sca}}  \int_{-1}^{1}  (|S_\parallel|^2 + |S_{\bot}|^2) \,u\, du,
\end{equation}
where $u=\mathrm{cos}(\theta)$. The asymmetry parameter is bounded by $[-1,1]$.  If the sphere scatters light equally between the forward and backward hemisphere, $g$ is zero. For forward scattering pattern, $g$ is positive. For spherical objects, the asymmetry parameter $g$ is independent of the polarization state. For the scattering patterns represented in Fig.~6, the asymmetry parameter values are around 0.85. Hence, the scattering power is concentrated in a solid angle of $35^{\circ}$.   

In conclusion,  we have shown that electromagnetically large nonmagnetic spheres made from low permittivity materials exhibit a notable number of closely packed multipole resonances where both the electric and magnetic multipolar moments have comparable complex amplitudes $a_\textrm{n} \approx b_\textrm{n}$ for all $n$. Therefore, the electromagnetic duality is approximately met in a broadband fashion even though the materials are nonmagnetic. This results in a broadband backscattering suppression accompanied with a considerable scattering efficiency.  In other words, the existence of significant higher order multipolar moments in these large spheres of low permittivity enhances the broadband effect, in stark contrast to high permittivity spheres, where terms higher than the dipoles typically degrade the ZBS performance. This result motivates further analysis on the use of dielectric spheres of low permittivity in all-dielectric metamaterials, for instance, to enhance the solar cell efficiency as an alternative to metallic nanoparticles that suffer from intrinsic losses \cite{akimov2010nanoparticle}. 
\section*{Methods}
The analysis is performed using a freely available Mie code\cite{matzler2002matlab} to compute the Mie coefficients of the sphere: $a_\textrm{n}$ and $b_\textrm{n}$. Using the Mie coefficients, the scattering efficiency $Q_\textrm{sca}$, the backscattering efficiency  $Q_\textrm{b}$, the  angular scattering parameters $S_\parallel$ and $S_{\bot}$, and the asymmetry parameter $g$ can be computed. The Wiscombe criterion has been used to guarantee convergence of all the shown quantities\cite{wiscombe1980improved, allardice2014convergence}. The analysis is done using lengths in units of the wavelength. This allows to easily scale the results to any desired spectral region as long as there are materials available with the required properties.


\begin{thebibliography}{1}

\expandafter\ifx\csname url\endcsname\relax
  \def\url#1{\texttt{#1}}\fi
\expandafter\ifx\csname urlprefix\endcsname\relax\def\urlprefix{URL }\fi
\providecommand{\bibinfo}[2]{#2}
\providecommand{\eprint}[2][]{\url{#2}}

\bibitem{mie1908beitrage}
\bibinfo{author}{Mie, G.}
\newblock \bibinfo{title}{Beitr{\"a}ge zur Optik tr{\"u}ber Medien, speziell
  kolloidaler Metall{\"o}sungen}.
\newblock \emph{\bibinfo{journal}{Annalen der physik}}
  \textbf{\bibinfo{volume}{330}}, \bibinfo{pages}{377--445}
  (\bibinfo{year}{1908}).

\bibitem{alu2008tuning}
\bibinfo{author}{Alu, A.} \& \bibinfo{author}{Engheta, N.}
\newblock \bibinfo{title}{Tuning the scattering response of optical
  nanoantennas with nanocircuit loads}.
\newblock \emph{\bibinfo{journal}{Nat. Photon.}}
  \textbf{\bibinfo{volume}{2}}, \bibinfo{pages}{307--310}
  (\bibinfo{year}{2008}).

\bibitem{knoll2000enhanced}
\bibinfo{author}{Knoll, B.} \& \bibinfo{author}{Keilmann, F.}
\newblock \bibinfo{title}{Enhanced dielectric contrast in scattering-type
  scanning near-field optical microscopy}.
\newblock \emph{\bibinfo{journal}{ Opt. Commun.}}
  \textbf{\bibinfo{volume}{182}}, \bibinfo{pages}{321--328}
  (\bibinfo{year}{2000}).

\bibitem{moreno2013analysis}
\bibinfo{author}{Moreno, F.}, \bibinfo{author}{Albella, P.} \&
  \bibinfo{author}{Nieto-Vesperinas, M.}
\newblock \bibinfo{title}{Analysis of the spectral behavior of localized
  plasmon resonances in the near-and far-field regimes}.
\newblock \emph{\bibinfo{journal}{Langmuir}} \textbf{\bibinfo{volume}{29}},
  \bibinfo{pages}{6715--6721} (\bibinfo{year}{2013}).

\bibitem{sarmecanic1997constraints}
\bibinfo{author}{Sarmecanic, J.}, \bibinfo{author}{Fomenkova, M.},
  \bibinfo{author}{Jones, B.} \& \bibinfo{author}{Lavezzi, T.}
\newblock \bibinfo{title}{Constraints on the nucleus and dust properties from
  mid-infrared imaging of comet hyakutake}.
\newblock \emph{\bibinfo{journal}{Astrophys. J. Lett.}}
  \textbf{\bibinfo{volume}{483}}, \bibinfo{pages}{L69} (\bibinfo{year}{1997}).

\bibitem{ferry2010design}
\bibinfo{author}{Ferry, V.~E.}, \bibinfo{author}{Munday, J.~N.} \&
  \bibinfo{author}{Atwater, H.~A.}
\newblock \bibinfo{title}{Design considerations for plasmonic photovoltaics}.
\newblock \emph{\bibinfo{journal}{Adv. Mater.}}
  \textbf{\bibinfo{volume}{22}}, \bibinfo{pages}{4794--4808}
  (\bibinfo{year}{2010}).

\bibitem{akimov2010nanoparticle}
\bibinfo{author}{Akimov, Y.~A.}, \bibinfo{author}{Koh, W.},
  \bibinfo{author}{Sian, S.} \& \bibinfo{author}{Ren, S.}
\newblock \bibinfo{title}{Nanoparticle-enhanced thin film solar cells: Metallic
  or dielectric nanoparticles?}
\newblock \emph{\bibinfo{journal}{Appl. Phys. Lett.}}
  \textbf{\bibinfo{volume}{96}}, \bibinfo{pages}{073111}
  (\bibinfo{year}{2010}).

\bibitem{zhang2015dielectric}
\bibinfo{author}{Zhang, Y.}, \bibinfo{author}{Nieto-Vesperinas, M.} \&
  \bibinfo{author}{S{\'a}enz, J.~J.}
\newblock \bibinfo{title}{Dielectric spheres with maximum forward scattering
  and zero backscattering: a search for their material composition}.
\newblock \emph{\bibinfo{journal}{J. Opt.}}
  \textbf{\bibinfo{volume}{17}}, \bibinfo{pages}{105612}
  (\bibinfo{year}{2015}).

\bibitem{chen2011optical}
\bibinfo{author}{Chen, J.}, \bibinfo{author}{Ng, J.}, \bibinfo{author}{Lin, Z.}
  \& \bibinfo{author}{Chan, C.}
\newblock \bibinfo{title}{Optical pulling force}.
\newblock \emph{\bibinfo{journal}{Nat. Photon.}}
  \textbf{\bibinfo{volume}{5}}, \bibinfo{pages}{531--534}
  (\bibinfo{year}{2011}).

\bibitem{kerker1983electromagnetic}
\bibinfo{author}{Kerker, M.}, \bibinfo{author}{Wang, D.-S.} \&
  \bibinfo{author}{Giles, C.}
\newblock \bibinfo{title}{Electromagnetic scattering by magnetic spheres}.
\newblock \emph{\bibinfo{journal}{J. Opt. Soc. Am.}} \textbf{\bibinfo{volume}{73}},
  \bibinfo{pages}{765--767} (\bibinfo{year}{1983}).



\bibitem{FerCor2012p}
\bibinfo{author}{Fernandez-Corbaton, I.} \emph{et~al.}
\newblock \bibinfo{title}{Electromagnetic duality symmetry and helicity
  conservation for the macroscopic maxwell's equations}.
\newblock \emph{\bibinfo{journal}{Phys. Rev. Lett.}}
  \textbf{\bibinfo{volume}{111}}, \bibinfo{pages}{060401}
  (\bibinfo{year}{2013}).

\bibitem{Zambrana2013}
\bibinfo{author}{Zambrana-Puyalto, X.}, \bibinfo{author}{Fernandez-Corbaton,
  I.}, \bibinfo{author}{Juan, M.~L.}, \bibinfo{author}{Vidal, X.} \&
  \bibinfo{author}{Molina-Terriza, G.}
\newblock \bibinfo{title}{Duality symmetry and kerker conditions}.
\newblock \emph{\bibinfo{journal}{Opt. Lett.}} \textbf{\bibinfo{volume}{38}},
  \bibinfo{pages}{1857--1859} (\bibinfo{year}{2013}).


\bibitem{FerCor2013c}
\bibinfo{author}{Fernandez-Corbaton, I.}
\newblock \bibinfo{title}{Forward and backward helicity scattering coefficients
  for systems with discrete rotational symmetry}.
\newblock \emph{\bibinfo{journal}{Opt. Express}} \textbf{\bibinfo{volume}{21}},
  \bibinfo{pages}{29885--29893} (\bibinfo{year}{2013}).

\bibitem{kuznetsov2011magnetic}
\bibinfo{author}{Kuznetsov, A.}, \bibinfo{author}{Miroshnichenko, A.},
  \bibinfo{author}{Fu, Y.}, \bibinfo{author}{Zhang, J.} \&
  \bibinfo{author}{Luk'yanchuk, B.}
\newblock \bibinfo{title}{Magnetic light.}
\newblock \emph{\bibinfo{journal}{Sci. Rep.}}
  \textbf{\bibinfo{volume}{2}}, \bibinfo{pages}{492}
  (\bibinfo{year}{2012}).

\bibitem{garcia2011strong}
\bibinfo{author}{Garc{\'\i}a-Etxarri, A.} \emph{et~al.}
\newblock \bibinfo{title}{Strong magnetic response of submicron silicon
  particles in the infrared}.
\newblock \emph{\bibinfo{journal}{Opt. Express}}
  \textbf{\bibinfo{volume}{19}}, \bibinfo{pages}{4815--4826}
  (\bibinfo{year}{2011}).

\bibitem{zambrana2013dual}
\bibinfo{author}{Zambrana-Puyalto, X.}, \bibinfo{author}{Vidal, X.},
  \bibinfo{author}{Juan, M.~L.} \& \bibinfo{author}{Molina-Terriza, G.}
\newblock \bibinfo{title}{Dual and anti-dual modes in dielectric spheres}.
\newblock \emph{\bibinfo{journal}{Opt. Express}}
  \textbf{\bibinfo{volume}{21}}, \bibinfo{pages}{17520--17530}
  (\bibinfo{year}{2013}).

\bibitem{geffrin2012magnetic}
\bibinfo{author}{Geffrin, J.-M.} \emph{et~al.}
\newblock \bibinfo{title}{Magnetic and electric coherence in forward-and
  back-scattered electromagnetic waves by a single dielectric subwavelength
  sphere}.
\newblock \emph{\bibinfo{journal}{Nat. Commun.}}
  \textbf{\bibinfo{volume}{3}}, \bibinfo{pages}{1171} (\bibinfo{year}{2012}).

\bibitem{fu2013directional}
\bibinfo{author}{Fu, Y.~H.}, \bibinfo{author}{Kuznetsov, A.~I.},
  \bibinfo{author}{Miroshnichenko, A.~E.}, \bibinfo{author}{Yu, Y.~F.} \&
  \bibinfo{author}{Luk'yanchuk, B.}
\newblock \bibinfo{title}{Directional visible light scattering by silicon
  nanoparticles}.
\newblock \emph{\bibinfo{journal}{Nat. Commun.}}
  \textbf{\bibinfo{volume}{4}},
  \bibinfo{pages}{1527}
   (\bibinfo{year}{2013}).

\bibitem{zhang2015colloidal}
\bibinfo{author}{Zhang, S.} \emph{et~al.}
\newblock \bibinfo{title}{Colloidal moderate-refractive-index cu2o nanospheres
  as visible-region nanoantennas with electromagnetic resonance and directional
  light-scattering properties}.
\newblock \emph{\bibinfo{journal}{Adv. Mater.}}
  \textbf{\bibinfo{volume}{27}}, \bibinfo{pages}{7432--7439}
  (\bibinfo{year}{2015}).

\bibitem{person2013demonstration}
\bibinfo{author}{Person, S.} \emph{et~al.}
\newblock \bibinfo{title}{Demonstration of zero optical backscattering from
  single nanoparticles}.
\newblock \emph{\bibinfo{journal}{Nano Lett.}} \textbf{\bibinfo{volume}{13}},
  \bibinfo{pages}{1806--1809} (\bibinfo{year}{2013}).

\bibitem{staude2013tailoring}
\bibinfo{author}{Staude, I.} \emph{et~al.}
\newblock \bibinfo{title}{Tailoring directional scattering through magnetic and
  electric resonances in subwavelength silicon nanodisks}.
\newblock \emph{\bibinfo{journal}{ACS Nano}} \textbf{\bibinfo{volume}{7}},
  \bibinfo{pages}{7824--7832} (\bibinfo{year}{2013}).

\bibitem{decker2015high}
\bibinfo{author}{Decker, M.} \emph{et~al.}
\newblock \bibinfo{title}{High-efficiency dielectric huygens surfaces}.
\newblock \emph{\bibinfo{journal}{Adv. Opt. Mater.}}
  \textbf{\bibinfo{volume}{3}}, \bibinfo{pages}{813--820}
  (\bibinfo{year}{2015}).

\bibitem{alaee2015generalized}
\bibinfo{author}{Alaee, R.}, \bibinfo{author}{Filter, R.},
  \bibinfo{author}{Lehr, D.}, \bibinfo{author}{Lederer, F.} \&
  \bibinfo{author}{Rockstuhl, C.}
\newblock \bibinfo{title}{A generalized kerker condition for highly directive
  nanoantennas}.
\newblock \emph{\bibinfo{journal}{Opt. Lett.}}
  \textbf{\bibinfo{volume}{40}}, \bibinfo{pages}{2645--2648}
  (\bibinfo{year}{2015}).

\bibitem{li2014broadband}
\bibinfo{author}{Li, Y.} \emph{et~al.}
\newblock \bibinfo{title}{Broadband zero-backward and near-zero-forward
  scattering by metallo-dielectric core-shell nanoparticles.}
\newblock \emph{\bibinfo{journal}{Sci. Rep.}}
  \textbf{\bibinfo{volume}{5}}, \bibinfo{pages}{12491}
  (\bibinfo{year}{2014}).

\bibitem{kerker2016scattering}
\bibinfo{author}{Kerker, M.}
\newblock \emph{\bibinfo{title}{The scattering of light and other
  electromagnetic radiation}} (\bibinfo{publisher}{Elsevier},
  \bibinfo{year}{2016}).


\bibitem{bohren1983absorption}
\bibinfo{author}{Bohren, C.~F.} \& \bibinfo{author}{Huffman, D.~R.}
\newblock \emph{\bibinfo{title}{Absorption and Scattering of Light by Small
  Particles}} (\bibinfo{publisher}{John Wiley \& Sons},
  \bibinfo{year}{2008}).





\bibitem{grainger2004calculation}
\bibinfo{author}{Grainger, R.~G.}, \bibinfo{author}{Lucas, J.},
  \bibinfo{author}{Thomas, G.~E.} \& \bibinfo{author}{Ewen, G.~B.}
\newblock \bibinfo{title}{Calculation of mie derivatives}.
\newblock \emph{\bibinfo{journal}{Appl. Opt.}}
  \textbf{\bibinfo{volume}{43}}, \bibinfo{pages}{5386--5393}
  (\bibinfo{year}{2004}).

\bibitem{abramowitz1964handbook}
\bibinfo{author}{Abramowitz, M.} \& \bibinfo{author}{Stegun, I.~A.}
\newblock \emph{\bibinfo{title}{Handbook of mathematical functions: with
  formulas, graphs, and mathematical tables}}, vol.~\bibinfo{volume}{55}
  (\bibinfo{publisher}{Courier Corporation}, \bibinfo{year}{1964}).

\bibitem{yang2008optical}
\bibinfo{author}{Yang, M.~K.}, \bibinfo{author}{French, R.~H.} \&
  \bibinfo{author}{Tokarsky, E.~W.}
\newblock \bibinfo{title}{Optical properties of teflon{\textregistered} af
  amorphous fluoropolymers}.
\newblock \emph{\bibinfo{journal}{J. Micro/Nanolith. MEMS MOEMS
}} \textbf{\bibinfo{volume}{7}}, \bibinfo{pages}{033010}
  (\bibinfo{year}{2008}).

\bibitem{FerCor2015}
\bibinfo{author}{Fernandez-Corbaton, I.}, \bibinfo{author}{Fruhnert, M.} \&
  \bibinfo{author}{Rockstuhl, C.}
\newblock \bibinfo{title}{Dual and chiral objects for optical activity in
  general scattering directions}.
\newblock \emph{\bibinfo{journal}{ACS Photonics}} \textbf{\bibinfo{volume}{2}},
  \bibinfo{pages}{376--384} (\bibinfo{year}{2015}).

\bibitem{matzler2002matlab}
\bibinfo{author}{M{\"a}tzler, C.}
\newblock \bibinfo{title}{Matlab functions for mie scattering and absorption,
  version 2}.
\newblock \emph{\bibinfo{journal}{IAP Res. Rep}} \textbf{\bibinfo{volume}{8}},
  \bibinfo{pages}{1--24} (\bibinfo{year}{2002}).

\bibitem{wiscombe1980improved}
\bibinfo{author}{Wiscombe, W.~J.}
\newblock \bibinfo{title}{Improved mie scattering algorithms}.
\newblock \emph{\bibinfo{journal}{Appl. Opt.}}
  \textbf{\bibinfo{volume}{19}}, \bibinfo{pages}{1505--1509}
  (\bibinfo{year}{1980}).

\bibitem{allardice2014convergence}
\bibinfo{author}{Allardice, J.~R.} \& \bibinfo{author}{Le~Ru, E.~C.}
\newblock \bibinfo{title}{Convergence of mie theory series: criteria for
  far-field and near-field properties}.
\newblock \emph{\bibinfo{journal}{Appl. Opt.}}
  \textbf{\bibinfo{volume}{53}}, \bibinfo{pages}{7224--7229}
  (\bibinfo{year}{2014}).

 \end{thebibliography}

\end{document}